\documentclass[superscriptaddress,twocolumn,amsmath,aps]{revtex4}
\usepackage{graphicx,color}
\usepackage{bm}
\usepackage[hypertex]{hyperref}

\newcommand{\be}{\begin{equation}}
\newcommand{\ee}{\end{equation}}
\newcommand{\bea}{\begin{eqnarray}}
\newcommand{\eea}{\end{eqnarray}}
\newcommand{\bsube}{\begin{subequations}}
\newcommand{\esube}{\end{subequations}}

\newcommand{\Eq}[1]{Eq.\,(\ref{#1})}

\newcommand{\dg}{\dagger}
\newcommand{\la}{\langle}
\newcommand{\ra}{\rangle}

\newcommand{\nl}{\nonumber \\}

\usepackage{amsfonts}
\usepackage{amsmath}
\usepackage{amssymb}
\usepackage{graphicx}
\usepackage{type1cm}        
\usepackage{times}          
\usepackage{bm}
\usepackage{color}                      

\newcommand{\beq}{\begin{equation}}
\newcommand{\eeq}{\end{equation}}
\newcommand{\beqn}{\begin{eqnarray}}
\newcommand{\eeqn}{\end{eqnarray}}
\newcommand{\bsub}{\begin{subequations}}
\newcommand{\esub}{\end{subequations}}

\begin{document}

\title{ Quantum Trajectory Approach to
 Molecular Dynamics Simulation with Surface Hopping }

\author{Wei Feng}
\affiliation{Department of Physics, Beijing Normal University,
Beijing 100875, China}
\author{Luting Xu}
\affiliation{Department of Physics, Beijing Normal University,
Beijing 100875, China}

\author{Xin-Qi Li}
\email{lixinqi@bnu.edu.cn}
\affiliation{Department of Physics, Beijing Normal University,
Beijing 100875, China}
\author{Weihai Fang}
\affiliation{College of Chemistry, Beijing Normal University,
Beijing 100875, China}

\date{\today}

\begin{abstract}
The powerful molecular dynamics (MD) simulation
is basically based on a picture that
the atoms experience classical-like trajectories
under the exertion of classical force field
determined by the quantum mechanically solved electronic state.
In this work we propose a quantum trajectory approach
to the MD simulation with surface hopping,
from an insight that an effective
``observation" is actually implied in the MD simulation
through tracking the forces experienced, just like checking
the meter's result in the quantum measurement process.
This treatment can build the nonadiabatic surface hopping
on a {\it dynamical} foundation, instead of the usual
artificial and conceptually inconsistent hopping algorithms.
The effects and advantages of the proposed scheme
are preliminarily illustrated by a two-surface model system.
\end{abstract}


\maketitle



A full quantum mechanical treatment for molecular dynamics (MD)
would break down along the increase of atomic
degrees of freedom. As a result, in practice
the MD technique based on an assumption
that atomic motions are governed by {\it classical} mechanics
has proved to be a very powerful tool
\cite{MD-1,MD-2,MD-4,Tul98a,Bar11}.
The MD simulations usually rely on the Born-Oppenheimer (BO)
approximation, which decouples the electronic and atomic motions.
That is, the atoms experience classical-like trajectories
under the exertion of appropriate classical force field,
which is determined by a single potential energy surface (PES)
associated with a single electronic state.
On the other aspect, the electronic states are solved
from the time-independent Schr\"odinger equation
for a series of atomic geometries (configurations).



In many situations, however, when the energy separation of
different PESs becomes comparable with the magnitude of the
nonadiabatic coupling (typically in the proximity of
conical intersection, see Fig.\ 1),
the BO approximation may often fail.
In this case, since each nondegenerate electronic state
defines a distinct BO PES, a transition between electronic states
would change, often drastically, the forces experienced by the atoms.
Proper account for this {\it nonadiabatic} effect
is of crucial importance in practice.
The treatment of nonadiabatic effects in MD has a long history.
The most widely applied include the so-called  ``Ehrenfest" or
``time-dependent-Hartree mean-field (TDHMF)" approach
\cite{Rat82,Mic83,Tul98a,Fri05},
the ``trajectory surface-hopping (TSH)¡± methods
\cite{Nik68,Tul71,Mi72,Kun79,Bla83,Mi83,Tul90,Tul94},
and their mixed schemes \cite{Kun91,Zhu04,Jan09}.
The former approach views that the electronic wavefunction
might in general be a linear combination of the BO adiabatic functions,
the atomic effective potential is thus calculated by
averaging the electronic Hamiltonian over such wavefunction,
while the BO adiabatic functions are
determined self-consistently with the atomic trajectory.
The TSH, a more appropriate methodology for dynamics
propagation of nonadiabatic systems,
is however based on an insight that
the trajectories should split into branches,
i.e., each trajectory should be on one state
or another, not somewhere in between.
And, the trajectories distribution is achieved by allowing hops
between surfaces according to some probability distribution.

%
%

\begin{figure}[!htbp]
  \centering
  \includegraphics[width=6.5cm]{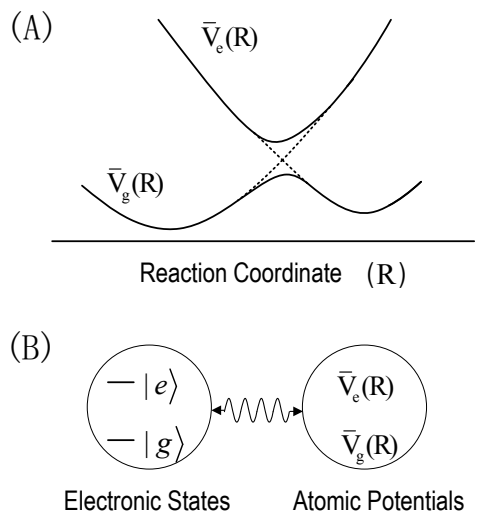}
  \caption{ (A) Schematic atomic potentials in terms of
adiabatic (solid) and diabatic (dashed) representations
for the electronic ground and first excited states
along the reaction coordinate.
(B) Measurement analogy by dividing the molecular system
into two subsystems,
and regarding the atomic part as a ``measurement apparatus"
which continuously probes the distinct electronic states
by the experienced distinct forces.    }
\end{figure}

A variety of trajectory-based methods have been developed.
Among them, the most typical one is Tully's
{\it fewest switches surface hopping} (FSSH) algorithm \cite{Tul90}.
This algorithm is based on the following insights/arguments:
{\it (i)}
The atomic trajectories determine the probabilities of
electronic transitions and the electronic transitions,
in turn, strongly influence the forces governing the trajectories.
{\it (ii)}
The atoms evolve on individual single PES and the nonadiabatic
effects are included by allowing hopping from one PES to another
{\it according to the weight of the respective electronic state}.
In the past two decades, stimulated by Tully's work,
a variety of variants of the FSSH algorithm
and a large number of applications have been carried out
\cite{Cok9192,Cok93,Tul9495,Ros97,Tul98,Fri91a,Fri94,Las08a,Las08b,Fab08,Hay09}.



Nevertheless, the {\it coherence} between states,
maintained in the FSSH algorithm
is usually wrong because of the independent trajectory approach
(see the recent review \cite{Bar11}).
The underlying {\it conceptual} difficulty can be further
understood as follows. In the FSSH algorithm,
the nonadiabatic effect is treated
by allowing hopping from one PES to another,
with the hopping probability determined by
the weight change of the respective electronic states
in {\it quantum superposition}.
Physically, however, the atomic {\it classical} trajectory
along a specific individual PES implies that it must
{\it collapse} the electronic state onto the corresponding BO
wavefunction, according to the {\it distinct} classical force experienced.
In other words, we can {\it no longer} treat the electronic state
in a quantum superposition of the many different components,
as in contrast proposed in the FSSH approach \cite{Tul90}.
As shown in Fig.\ 1(B), through an analogy with quantum measurement
or with the more popular Schr\"odinger's Cat paradox,
the FSSH algorithm simply means that we are still
treating the radioactive decay in a {\it quantum superposition state},
while having found the Cat definitely ``alive" or ``dead".

In this work, rather than designing certain
wiser algorithms for the TSH probability,
we would like to {\it view} the problem {\it in an alternative way}
and accordingly propose a quantum trajectory approach to the MD simulation.
That is, we are able to develop a {\it self-consistent} in physics
and hopefully highly implementable stochastic MD scheme,
which quite naturally renders the TSH in the most fundamental notion
of {\it quantum jump}.


\vspace{0.5cm}
{\it Formulation.}---
For the purpose to be clear soon,
we introduce the electronic Hamiltonian
by subtracting the atomic kinetic energy
from the total Hamiltonian as
\bea
H_{el}(r,\textbf{R}) = -\sum_j \frac{\hbar^2}{2m_j}\nabla_j^2
   + V(r,\textbf{R}) .
\eea
Here, the first term describes the kinetic energy of electrons.
We use $r$ denoting all the electronic coordinates,
i.e., $r\equiv \{r_j; j=1,2,\cdots \}$.
And, formally, $\textbf{R}$ denotes the atomic configuration.

We expand the electronic wavefunction by a set of known
basis functions,
$\Psi(r,\textbf{R})=\sum_jc_j(t)\phi_j(r,\textbf{R})$.
In principle, $\{ \phi_j(r,\textbf{R}) \}$ can be any
electronic basis functions,
while in practice it would be convenient to chose them
as the Born-Oppenheimer adiabatic wavefunctions.
That is, $H_{el}(r,\textbf{R})\phi_j(r,\textbf{R})
= \varepsilon_j(\textbf{R}) \phi_j(r,\textbf{R})$,
which means that $\phi_j(r,\textbf{R})$ are the instantaneous
eigenstates of $H_{el}(r,\textbf{R})$, for a given
$\textbf{R}\equiv\textbf{R}(t)$.
These basis wavefunctions, together with the corresponding
eigen-energies $\varepsilon_j(\textbf{R})$, are the standard
output of the usual quantum chemistry computation.
Using the selected basis functions, we further introduce
\bea
V_{jk}(\textbf{R})&=&\la \phi_j(r,\textbf{R}) | H_{el}(r,\phi_j(r,\textbf{R}))
 |\phi_k(r,\textbf{R})\ra ,  \\
\textbf{d}_{jk}(\textbf{R})&=& \la \phi_j(r,\textbf{R})|
\nabla_{\textbf{R}} |\phi_k(r,\textbf{R})\ra .
\eea
Obviously, for the BO adiabatic basis,
$V_{jk}(\textbf{R})$ is diagonal,
$V_{jk}(\textbf{R})=\varepsilon_j(\textbf{R})\delta_{jk}
\equiv \bar{V}_j(\textbf{R})\delta_{jk}$.
Each $\bar{V}_j(\textbf{R})$ is known as
the BO potential energy surface (PES).
The quantity $\textbf{d}_{jk}(\textbf{R})$
characterizes {\it nonadiabatic coupling}
between the ($j$th and $k$th) PESs,
which has an effect of driving surface hopping.
This can be seen more clearly by the following.
The {\it coherent} electronic evolution is governed
by the Schr\"odinger equation $i\hbar\dot{\Psi}=H_{el}\Psi$,
which results in a set of coupled equations for $c_j(t)$
\bea\label{Seq-1}
i\hbar \dot{c}_j &=& \sum_k \left[ V_{jk}(\textbf{R})
  -i\hbar \dot{\textbf{R}}
  \cdot \textbf{d}_{jk}(\textbf{R}) \right]c_k  \nl
 &\equiv & \sum_{jk}\tilde{H}_{jk}(\textbf{R}) c_k .
\eea
In this expression, we can understand $\tilde{H}_{jk}(\textbf{R})$
as an effective Hamiltonian matrix in the selected basis,
which straightforwardly describes the evolution of
the state amplitudes $c_j(t)$.
In terms of a projective operator form,
the effective Hamiltonian can be expressed as
$\tilde{H}(\textbf{R})=\sum_{jk} \tilde{H}_{jk}(\textbf{R})
|\phi_j(\textbf{R})\ra  \la \phi_k(\textbf{R})|$.
Using the property
$\textbf{d}^*_{jk}(\textbf{R})=-\textbf{d}_{kj}(\textbf{R})$,
one can prove that $\tilde{H}(\textbf{R})$ is Hermitian,
as it should be.
Moreover, corresponding to the BO adiabatic basis,
we have even simpler results:
$\tilde{H}_{jj}(\textbf{R})\equiv \varepsilon_j(\textbf{R})$;
and
$\tilde{H}_{jk}(\textbf{R})\equiv \Omega_{jk}(\textbf{R})
=-i\hbar \dot{\textbf{R}} \cdot \textbf{d}_{jk}(\textbf{R})$,
if $j\neq k$. In this way, we see that it is the product of
$\dot{\textbf{R}}$ and $\textbf{d}_{jk}(\textbf{R})$ that characterizes
the nonadiabatic coupling between the BO potential surfaces.


It was merely based on \Eq{Seq-1}, which describes the {\it quantum mechanical}
evolution of the superposition amplitudes of electronic states,
that the highly celebrated FSSH algorithm was proposed \cite{Tul90},
with the following central idea.
From the ``normalized" population change of the electronic state
$|\phi_j(\textbf{R})\ra$,
defined as $p_j(t)=[|c_j(t)|^2-|c_j(t+\Delta t)|^2]/|c_j(t)|^2$,
one performs a standard Monte-Carlo choice (with this probability
$p_j(t)$) to determine whether or not a {\it hopping event} should
take place, from the BO potential surface
$\bar{V}_j(\textbf{R})$ to another one.

Intuitively, this looks indeed a promising algorithm,
since the BO potential surface $\bar{V}_j(\textbf{R})$
is anyhow associated with the electronic state $|\phi_j(\textbf{R})\ra$,
one can then conclude that the {\it change} of the electronic occupation
probability {\it must} imply a surface-hopping event occurred.
Nevertheless, in each single realization of MD trajectory,
the atomic motion experiences a series of distinct PESs
(but not their weighted superposition),
together with successive hopping between them.
Based on the entanglement-type correlation between
the electronic states and the atomic PESs, see Fig.\ 1(B),
the {\it classical trajectory treatment} for the atomic motion
must {\it collapse} the electronic state to an individual one,
with the result determined by the classical force
experienced by the atoms.
In other words, we can no longer {\it inappropriately} maintain
the electronic state in a quantum superposition of two (or many)
different components $|\phi_j(\textbf{R})\ra$,
as in contrast assumed in the FSSH approach \cite{Tul90}.
In essence, using the term of Schr\"odinger's Cat paradox,
the mixed quantum-classical FSSH treatment
is {\it equivalent} to treating the radioactive decay still
in ``quantum superposition",
while one has definitely found the Cat ``alive" or ``dead".


In order to eliminate the above {\it inconsistency}
in the FSSH algorithm, we propose that one should apply
a measurement-based description to the electronic state evolution,
instead of using the $\textbf{R}$-dependent Schr\"odinger equation.
Indeed, since $\textbf{R}$ enters the Schr\"odinger equation,
this partly accounts for the effect of the atomic motion
on the electronic state evolution. {\it But this is not enough}.
Actually, the atomic motion is continuously getting
the {\it state information} of the electrons,
via the correlation between the PES and the electronic state.
As a result, in addition to involving ``$\textbf{R}$" as an
external parameter in the electronic Schr\"odinger equation,
one should at the same time account for the {\it backaction}
of the electronic-state {\it information gain}.
This means that the atomic motion is {\it in essence}
making a continuous quantum measurement on the electronic state.
The distinct ``force" experienced by the atomic trajectory motion,
effectively, plays the same role
as the meter's {\it output} in quantum measurement process.

After the above {\it conceptual} preparation, we can then apply
the established quantum trajectory equation (QTE)
to account for the backaction effect owing to
{\it state information gain} as follows \cite{WM10}
\bea\label{QTE}
\dot{\rho}&=&-\frac{i}{\hbar}  \left[\tilde{H}(\textbf{R}),\rho\right]
+\sum_j \Gamma_{\phi j}
 {\cal D}\left[M_j(\textbf{R})\right]\rho \nl
&& + \sum_j \sqrt{\Gamma_{\phi j}}
{\cal H}\left[M_j(\textbf{R})\right]\rho~\xi_j(t) .
\eea
Here, $\rho$ is the electronic state density matrix,
with elements $\rho_{jk}$. The diagonal elements describe
population probabilities on the BO states,
while the off-diagonal elements describe quantum coherence
between them.
Note that, on the left hand side of \Eq{QTE},
we have made a convention that $\dot{\rho}$ represents
$\sum_{jk} \dot{\rho}_{jk} |\phi_j(\textbf{R})\ra
\la \phi_k(\textbf{R})|$,
but not
$\frac{d}{dt}[\sum_{jk}\rho_{jk} |\phi_j(\textbf{R})\ra
\la \phi_k(\textbf{R})|]$.
Accordingly, $\tilde{H}(\textbf{R})$ in the above QTE
is the effective Hamiltonian defined in \Eq{Seq-1}.


In \Eq{QTE}, the measurement operator,
$M_j(\textbf{R})=|\phi_j(\textbf{R})\ra  \la \phi_j(\textbf{R})|$,
corresponds to an identification (discrimination) to the electronic
state $\phi_j(\textbf{R})$ by the atomic motion.
It is well known that any state {\it discrimination}
will destroy the quantum coherence (the quantum superposition).
This effect is described by the second term in \Eq{QTE},
with the dephasing rate $\Gamma_{\phi j}$
corresponding to discrimination of the state $\phi_j(\textbf{R})$.
Here the Lindblad-type superoperator means,
${\cal D}[M_j]\rho = M_j\rho M_j^{\dg}
- \frac{1}{2}\{ M_j^{\dg}M_j, \rho\} $.
Finally and {\it very importantly}, the last term in \Eq{QTE}
accounts for the effect of state {\it information gain}.
The involved superoperator, more explicitly, is defined as
${\cal H}[M_j]\rho = M_j\rho + \rho M_j^{\dg}
- \la M_j+M_j^{\dg} \ra \rho $, where
$\la M_j+M_j^{\dg}\ra\equiv {\rm Tr}[(M_j+M_j^{\dg})\rho]$.
$\xi_j(t)$ are the Gaussian stochastic noises,
satisfying the ensemble average property
${\rm E}[\xi_j(t)\xi_k(t')]=\delta_{jk}\delta(t-t')$.
Notice that the last term in \Eq{QTE} is not originated from
any {\it external} noise, but from an {\it intrinsic}
stochastic collapse (quantum jump)
associated with ``observation",
i.e., the quantum measurement postulate.

Applying \Eq{QTE} in a strong observation regime
(with large $\Gamma_{\phi j}$), one can generate
the desired TSH picture.
That is, the large $\Gamma_{\phi j}$ correspond to
MD simulation under distinct forces.
Then, under the influence of the atomic trajectory motion,
the nonadiabatic coupling will drive a series of
stochastic {\it jumps} between the BO states.
This is nothing but the desired ``surface hopping" behavior
which can be generated, in the existing schemes, only by
constructing various hopping ``algorithms", non-dynamically.


Alternatively, applying \Eq{QTE} in a weaker observation regime
(with smaller $\Gamma_{\phi j}$),
while the atomic motion propagates along a single PES
away from the level-crossing area,
the Ehrenfest-type mean field approach is
restored in the proximity of the level-crossing area.
That is, by computing the potential energy via
$\bar{V}(\textbf{R})={\rm Tr}[H_{el}(r,\textbf{R})\rho(t)]$,
the effect of quantum superposition of the BO potential surfaces
is taken into account.
In our opinion, this effect should exist there.
Using the term of quantum measurement, this simply corresponds to
an {\it incomplete observation} in the level crossing area
where the atomic motion,
particularly in the case of strong nonadiabatic coupling,
does not {\it fully} distinguish {\it which} potential surface,
but only experiences a mixed force.
We believe that applying the proposed approach
to the above regime is a valuable extension
of the hybrid TSH-TDHMF scheme.


\vspace{0.5cm}
{\it Illustrative Demonstration.}---
Below we illustrate how the approach proposed above
can generate the TSH behavior, {\it dynamically}
(or {\it physically} in essence),
instead of using the {\it probability algorithms}
to generate the ``hopping" events.
For simplicity, we restrict our demonstration to
a one-dimensional (1D) two-surface example.
Because of the 1D nature, in what follows we
denote ``$\textbf{R}$" simply by ``$R$".
To be specific, we model the two BO potential surfaces
as schematically shown in Fig.\ 2(A).
For $|R-R_0|>a$,
we assume $\varepsilon_2(R)-\varepsilon_1(R)
=\alpha |R-R_0|$;
while for $|R-R_0|\leq a$,
the adiabatic energy difference is a constant,
$\varepsilon_2(R)-\varepsilon_1(R)=\Delta$.
Here, the scale of the level-crossing area
is determined by $a=\Delta/\alpha$.
Rather than a real MD simulation, following the Landau-Zener model,
we assume a constant speed for the atomic motion, i.e., $R(t)=v t$.
Moreover, the nonadiabatic coupling is parameterized as
$\tilde{H}_{12}=(\tilde{H}_{21})^*=-i\hbar v\beta\equiv -i\hbar\Omega$,
from an assumption of $\dot{R}\cdot d_{12}(R)=v\beta$.

\begin{figure}[!htbp]
  \centering
  \includegraphics[width=8.0cm]{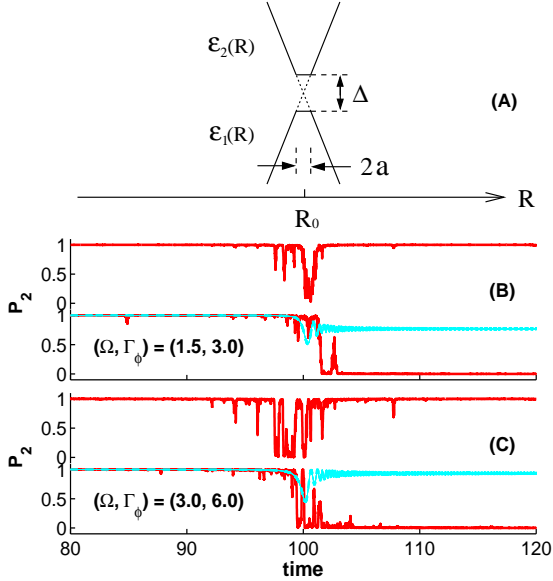}
  \caption{ Demonstration of the surface-hopping behavior.
 (A): The model we used in our numerical simulation.
 (B) and (C): Four representative trajectories (red curves) from
 two groups of parameters, while the cyan curves stand for the results
 of $\Gamma_{\phi}=0$ in the absence of ``observation".
 Other parameters: $\Delta=1.0$ and $\alpha=10$.
 In the simulation (also in Fig.\ 3 and in the main-text discussion)
 we assume $\Gamma_{\phi 1}=\Gamma_{\phi 2}\equiv\Gamma_{\phi}$.  }
\end{figure}

In Fig.\ 2(B) and (C) we display some representative
quantum trajectories. First, we notice that
the nonadiabatic coupling (with {\it constant} magnitude ``$\Omega$")
can cause efficient transitions {\it largely}
in the proximity of the conical intersection area, because of
the relatively small energy separation between the PESs.
Also, in the whole range, \Eq{QTE} will give a quantum mechanically
{\it pure} (but not {\it unitary}) evolution for the electronic state.
Fig.\ 2(B) corresponds to a situation with
relatively weak coupling and weak observation.
In this case, in the intersection area
the electronic state consists of a superposition of the BO basis states.
Using this state to calculate the atomic forces,
desirably, should be an extension of the usual mean-field approach.
However, if we increase $\Gamma_{\phi}$ further, as shown in Fig.\ 2(C),
in the proximity of the central level-crossing area
the sharp ``surface hopping" behavior will appear eventually.

We emphasize in this context that, very importantly,
even for a weak strength $\Gamma_{\phi}$
the trajectory will {\it gradually} and {\it stochastically}
collapse onto one of the energy surfaces.
This is in sharp contrast with the prediction
of the Schr\"odinger equation.
In that case, the state will remain in quantum superposition,
after the evolution passes through the central intersection area.
We demonstrate this feature by the cyan curves in both Fig.\ 2(B) and (C),
where we see that the final state is indeed in superposition.
However, a continuous ``observation" will collapse the state, gradually,
onto one of the basis states, either $|\phi_1\ra$ or $|\phi_2\ra$.
Then the desired picture of trajectory propagation along a single energy
surface is generated, by this dynamical and physical means.


In practice, to compare the stochastic MD simulation with experiments,
one should ensemble-average the stochastic trajectories.
For the assumed {\it constant} speed of atomic motion,
the trajectory distribution can be simply predicted
by averaging the QTE first.
This can be done easily by removing the last (unraveling)
term in \Eq{QTE} and remaining only the first two terms.
We actually obtain the usual master equation.
Then, from the density matrix, we get the probabilities
$p_1(t)=\rho_{11}(t)$ and $p_2(t)=\rho_{22}(t)$,
which give the distribution ratio of the atomic trajectories
in the asymptotic limit ($t\rightarrow\infty$).
We should note that, however, in practical MD simulations
one {\it cannot} average first \Eq{QTE},
since for different trajectories the stochastic
atomic forces experienced would be different.
Obviously, this feature cannot be captured
by the averaged master equation.
In other words,
the averaged master equation does not allow us
to correctly compute the atomic forces, we are
then unable to propagate the atomic motion based on Newton's law.
This is the reason that in literature, when one
attempts to introduce {\it decoherence},
the master equation approach does not work.

\begin{figure}[!htbp]
  \centering
  \includegraphics[width=7.0cm]{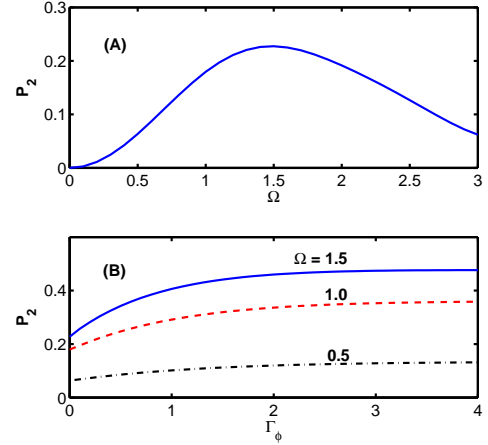}
  \caption{ Nonadiabatic transition probability from the initial state
  $|\phi_1(R_{-\infty})\ra$ to the final one $|\phi_2(R_{\infty})\ra$.
  $R_{\mp\infty}$ represent the initial and final asymptotic
  reaction coordinates. This probability indicates the distribution
  of the final ``products".
  (A) Dependence on the nonadiabatic coupling strength
  in the absence of dephasing ($\Gamma_{\phi}=0$).
  (B) Dephasing effect on the transition probability.
  Parameters: $\Delta=1.0$ and $\alpha=10$.}
\end{figure}

In Fig.\ 3(A), we show the dependence of the nonadiabatic
transition on the coupling strength. We take $\Gamma_{\phi}=0$
in order to reveal this mere dependence more clearly.
We observe a {\it turnover} behavior,
with an optimal nonadiabatic coupling
to make the transition reach maximum.
This feature indicates, in some counterintuitive manner,
that a faster atomic motion may not necessarily enhance
the transition probability.
This behavior differs somehow from the prediction
of the Landau-Zener formula, since that formula will
give a larger transition probability for a faster speed.
The reason is that, here, in the adiabatic BO basis,
the faster atomic motion will enhance
the nonadiabatic coupling between the energy surfaces,
while in the Landau-Zener model,
expressed in a diabatic basis, the passage speed across
the intersection area has no such effect .


In Fig.\ 3(B), we show the decoherence effect
on the nonadiabatic transition.
Here we observe a {\it decoherence-enhanced} transition behavior.
But, this feature is not universal.
It only corresponds to a {\it weak} dynamic tunneling regime.
Actually, not shown in Fig.\ 3(B), if we tune the coupling strength
into an intermediate tunneling regime, a {\it turnover} behavior
can appear, whereas a {\it decoherence-suppressed} transition will
take place in strong tunneling regime.
We also observe that, if an efficient coupling persists
in the intersection area for longer time, the decoherence will
result in a final {\it equal occupation} on the both energy surfaces,
as shown by the curve with $\Omega=1.5$ in Fig.\ 3(B).


Finally, we remark that a {\it decoherence} in the TSH should arise
because the MD simulation is propagated along a single trajectory
determined by the gradients for an specific electronic state.
In this propagation, the amplitudes of all other states
are {\it artificially} restricted to be also propagated along the same
trajectory, even though the gradients of these other states
would dissipate these amplitudes along other directions.
This effect has been discussed in detail
and corrections to it have been proposed
\cite{Sch96,Tha98,Zhu05,Gra07,Gra10}.
For instance, in Ref.\ \cite{Gra07}, it was shown that the
divergence between occupation and average population
is caused by the missing decoherence, and that this discrepancy
can be eliminated if the decoherence corrections are incorporated.

\vspace{0.5cm}
{\it Acknowledgments.}---
We thank Jiushu Shao, Zhenggang Lan and Sheng Meng
for stimulating discussions.
This work was supported by the Major State Basic Research
Project of China (No.\ 2011CB808502 \& 2012CB932704)
and the NNSF of China (No. 101202101 \& 10874176).

\appendix
\section{Measurement Interpretation }

In this appendix we present a further discussion to relate
the stochastic MD simulation with quantum measurement.
As mentioned in the main text,
we can divide the whole system into two coupled
electronic and atomic {\it subsystems}, as shown in Fig.\ 1(B).
Under the BO approximation, the relatively {\it heavy} and {\it slowly moving}
atomic subsystem largely experiences a classical-like trajectory motion.
Therefore, the atomic subsystem looks very like a measurement meter,
which is continuously ``measuring" the electronic states
by its experienced distinct forces.
In this context, the PES (or force) is the meter's output
(measurement result), from which the electronic state is inferred.

In a more heuristic manner, we may
relate the problem to the paradox of Schr\"odinger's Cat.
As illustrated in Fig.\ 1(B),
we assume two electronic states, $|g\ra$ and $|e\ra$.
Accordingly, we have two PESs, $\bar{V}_g(R)$ and $\bar{V}_e(R)$,
which correspond to the Cat states ``alive" and ``dead".
If we insist on regarding the {\it whole system} as a {\it closed} one,
we will arrive to a (quantum) superposition of
$\bar{V}_g(R)$ and $\bar{V}_e(R)$, exactly as
the superposition of ``alive" and ``dead" states in the Cat paradox.
But, as is well known, the Cat paradox can be resolved {\it only}
by a neglect of the large number of microscopic
degrees of freedom of the Cat and the surrounding environment.
It is merely this treatment that can result in the emergence
of {\it classicality} which shows the Cat in a state of
either ``alive" or ``dead", but not in between.
We then arrive to two statements:
{\it (i)}
In the semi-classical stochastic MD simulation,
the essential picture of classical trajectory and surface hopping
has implied an incorporation of certain external environment
and an {\it effective} observation.
For a {\it closed} quantum system, there is no way
to generate the {\it jump} ({\it hopping}) behavior.
{\it (ii)}
We hypothesize that the TSH-based MD simulations
are more appropriate for many-atom system in condensed phase.
For an ``isolated (clean)" few-atom system,
a full quantum mechanical simulation should be required.

In the TSH-MD simulations, we can imagine that the ``observation"
is realized through tracking the forces experienced,
just like checking the meter's result
in the quantum measurement process.
Together with the quantum nonadiabatic coupling, successive jumps
then appear, exactly like in the continuous measurement
of a driving system\cite{WM10}.
Moreover, this approach is quite versatile.
For instance, it can describe the possible ``partial" hopping behavior.
That is, when the atomic motion passes through the conical intersection
region, the atoms may not be able to resolve the experienced force
from which PES.
In this case, the resultant state is still in quantum superposition.
However, the state evolution should obey the quantum
trajectory equation, \Eq{QTE}, but not the Schr\"odinger equation.
We believe that this treatment is a valuable extension
of the hybrid TSH-TDHMF approach\cite{Kun91,Zhu04,Jan09}.


\end{document}